\documentclass[pre,twocolumn,floatfix,superscriptaddress,showpacs,showkeys,preprintnumbers,amsmath,amssymb]{revtex4}

\usepackage{graphicx}
\usepackage{dcolumn}
\usepackage{bm}

\begin{document}

\preprint{APS/123-QED}

\title{Bifurcation and Chaos in Coupled Periodically Forced\\ 
Non-identical Duffing Oscillators} 
\author{U.~E.~Vincent}
\affiliation{Department of Physics, Olabisi Onabanjo University, Ago-Iwoye, Nigeria.}
\author{A. Kenfack}
\email[Corresponding author:]{kenfack@mpipks-dresden.mpg.de}
\affiliation{Max Planck Institute for the Physics of Complex Systems, N$\ddot
  o$thnitzer Strasse 38, 01187 Dresden, Germany.}

\date{\today}

\begin{abstract}
We study the bifurcations and the chaotic behaviour of a
periodically forced double-well Duffing oscillator coupled to a single-well
Duffing oscillator. Using the amplitude and the frequency of the driving force as
control parameters, we show that our model presents phenomena which were not observed in coupled periodically forced Duffing oscillators with identical potentials. In the regime of relatively weak coupling, bubbles of bifurcations and chains of symmetry-breaking are identified. For much stronger couplings, Hopf bifurcations born from orbits of higher periodicity and  supercritical Neimark bifurcations emerge. Moreover, tori-breakdown route to a strange non-chaotic attractor is also another highlight of features found in this model.
\end{abstract}

\maketitle
\section{Introduction}
The dynamics of coupled nonlinear oscillators has attracted considerable
attention in recent years because they arise in many branches of science. Coupled
oscillators are used in the modeling of many physical, chemical, biological
and physiological systems such as coupled p-n junctions \cite{buskirk1985}, charge-density waves \cite{strogatz1989}, chemical-reaction systems \cite{kuramoto1984},
and biological-oscillation systems \cite{winfree1980}. They exhibit varieties of bifurcations
and chaos~[5-12], pattern formation \cite{perez-Munuzuri1995}, synchronization~[7,14-20] and so on.

Bifurcation analysis is a useful and widely studied sub-field of dynamical systems.
The observation of the bifurcation scenario allows one to draw qualitative and quantitative conclusions about the structure and dynamics of the systems theory. 
Numerous theoretical and numerical studies
in this direction have been carried out for several specific problems [6-12]. Here,
we have referred to only recent studies relevant with coupled nonlinear systems. Of particular 
interest in this study are coupled Duffing oscillators - strictly
dissipative nonlinear oscillators which model the motion of various physical
systems such as a pendulum, an electrical circuit or a Josephson-junction to mention but a few. Two or more
identical or non-identical oscillators may be coupled in different ways. Thus, the type of the oscillators and couplings as well as the external perturbation have a significant influence on the dynamics of the coupled
system. For instance, Kozlowski et al. \cite{kozlowski1995} have shown that the global
pattern of bifurcation curves in the parameter space of two coupled identical periodically
driven single-well Duffing oscillators consists of repeated subpatterns of Hopf
bifurcations - a scenario that differs from that of a single Duffing oscillator.

Very recently, Kenfack~\cite{kenfack2003} studied the model already considered in~\cite{kozlowski1995}
but using  identical double-well potentials. That model presented many striking
departures from the behaviour of coupled single-well Duffing oscillators. Scenarios such as  multiple period-doubling of both types, symmetry-breaking, sudden chaos and a great abundance of Hopf bifurcations were found in that model, in addition to the well-known routes to chaos in a one-dimensional Duffing oscillator~\cite{parlitz1993}.

The present study derives its motivation from references~\cite{kozlowski1995,kenfack2003}. Here, we present a model 
consisting of a periodically driven double-well Duffing oscillator coupled to a single-well
Duffing oscillator. We numerically analyse varieties of bifurcations with special emphasis on those which are typical of the model. This is explored in different coupling regimes as a function of the two control parameters of the system namely, the amplitude $f$ and the angular frequency $\omega$ of the driving force.  

The paper is organized as follows: Section II describes the system under consideration. In
section III, we present several numerical results of bifurcation structures and
conclude the paper in section IV.
\section{The Model}
        The Duffing oscillator \cite{stoker1950} that we treat here is a well-known model
        of nonlinear oscillator. It is governed by the following dimensionless second-order differential equation
\begin{equation}
        \frac{d^2x}{dt^2} + b\frac{dx}{dt} + \frac{dV(x)}{dx} =f\cos (\omega t), \label{duffing}
\end{equation}
        where $x$ and $b$ stand for the position coordinate of a particle and
        the damping parameter, respectively. The right hand side of eq.~(\ref{duffing}), represents the driving force at  time t, with the amplitude $f$ and the angular frequency $\omega$.
        The oscillator belongs to the category of three-dimensional dynamical
        systems ($x,dx/dt,t$). The system (\ref{duffing}) is a generalization of the classic
        Duffing oscillator equation and can be considered in three main physical
        situations, wherein the dimensionless potential
\begin{equation}
        V(x) = \alpha \frac{x^{2}}{2} + \beta \frac{x^{4}}{4}, \label{potential1d}
\end{equation}
        is a (i) single-well ($\alpha>0, \beta>0$), (ii) double-well ($\alpha<0, \beta>0$)
        or (iii) double-hump ($\alpha>0, \beta<0$) potential. Each of the above three
        cases has become a classic central model describing inherently nonlinear
        phenomena exhibiting rich and baffling varieties of regular and chaotic
        motions. 

\begin{figure}[h]
\vskip 0.25cm
\includegraphics[width=7cm,height=14cm]{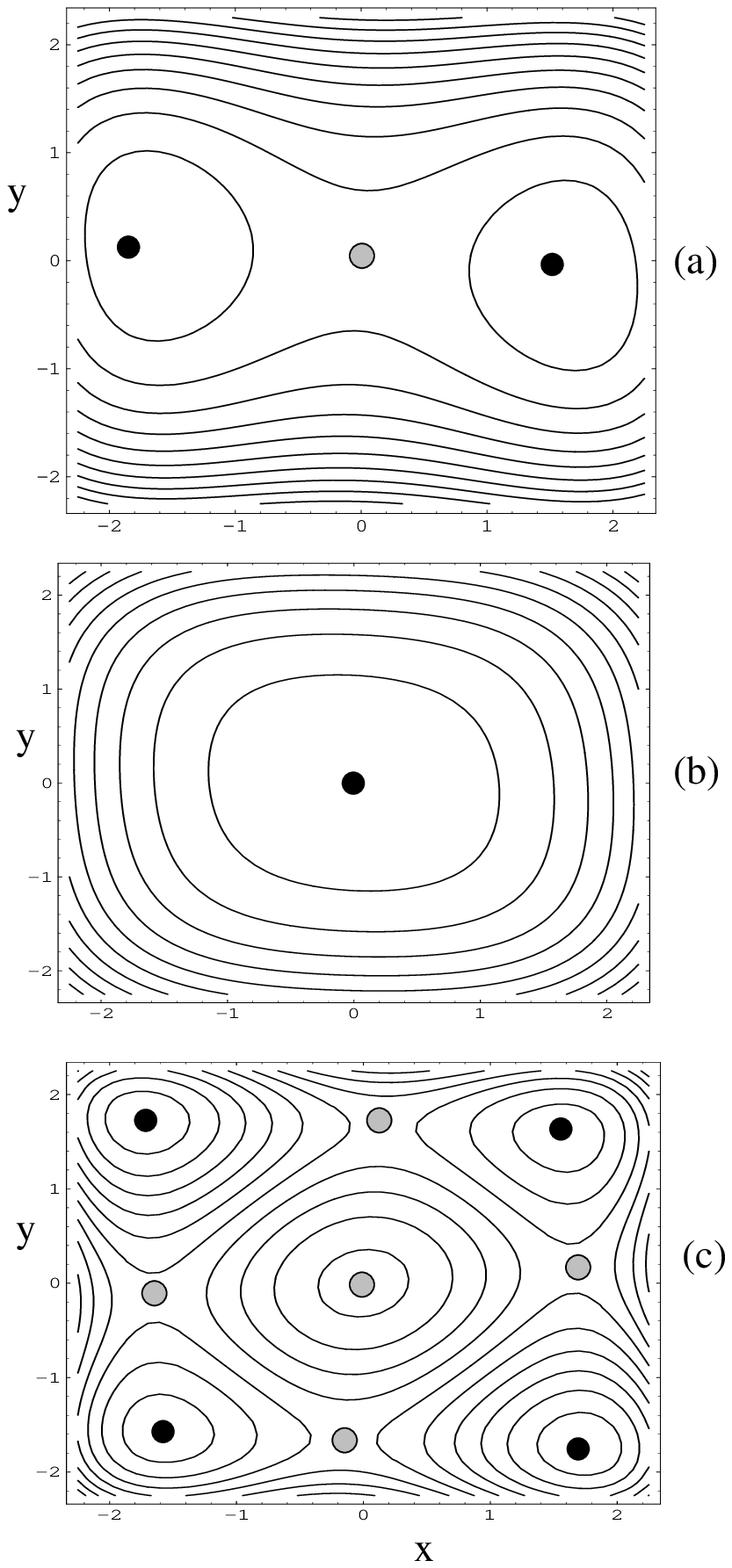}
\caption{\label{figure1} Contour plots of coupled Duffing oscillators for $k=0.1$ : (a) double-single wells, (b) single-single wells,
(c) double-double wells.}
\end{figure}

When two of such system eq.(~\ref{potential1d})  interact with each
        other through a specific coupling, the dynamics is expected to be even
        richer and more attractive. The coupling employed in the present paper
        is a linear feed-back which can be interpreted as a perturbation of each
        ocillator by a signal proportional to the differences of their
        positions. The potential governing such a coupled system reads
\begin{eqnarray}
V(x,y)=\frac{\alpha}{2}x^2+\frac{\beta}{4}x^4+\frac{\alpha'}{2}y^2+\frac{\beta'}{4}y^4+\frac{k}{2}(x-y)^2, \label{potential2d}
\end{eqnarray}
where $k$ is the coupling parameter. Depending on the values taken by the parameters
$\alpha,\beta,\alpha'$ and $\beta'$, one distinguishes three categories of globally
bounded coupled Duffing oscillators; namely the single-single wells, the single-double
wells and the double-double wells. Fig.~\ref{figure1} displays a prototypical
contour plot of the potential~(\ref{potential2d}) for a weak coupling strength
$k=0.1$. Grey dots denote unstable fixed points (saddles), while dark ones represent
stable fixed  points of the system. Clearly the single-double wells (Fig.~\ref{figure1}.a)
has an unstable fixed point and two stable fixed points; the single-single wells
(Fig.~\ref{figure1}.b) shows only one stable fixed point; however the double-double
wells (Fig.~\ref{figure1}.c) possesses five unstable fixed points and four stable
points.  When the coupling strenght $k$ increases, the positions of these fixed
points are modified accordingly and provide a variety of interesting phenomena
which we will study  in the subsequent sections. From eq.(~\ref{potential2d}), the equations of motion of the driven, coupled double-single wells ($\alpha=-\alpha'>0$,$\beta=\beta'>0$) Duffing oscillators corresponding to (Fig.~\ref{figure1}(a)) are derived and can be expressed as 
\begin{eqnarray}
        \frac{d^2x}{dt^2} &=& -b\frac{dx}{dt} +\alpha x-\beta x^{3}+k(y-x)+f\cos (\omega t)  \nonumber \\
      \frac{d^2y}{dt^2} &=& -b\frac{dy}{dt} -\alpha y-\beta y^{3}-k(y-x).  \label{equationsmotion}
\end{eqnarray}

       In the special case where $k=0$, it is obvious that system~(\ref{equationsmotion})
       reduces to two independent subsystems: the first being a periodically
       forced double-well Duffing oscillator with state variables
        ($x,dx/dt$), while the second is the unforced single-well Duffing
        oscillator having the state variables  ($y,dy/dt$). The extended phase
        space of our model is five-dimensional ($ \mathbb{R}^{4}\times S^{1}$), wherein any element of the state
        space is denoted by ($x,dx/dt,y,dy/dt,\theta$); $S^{1}$ being the
        unit circle containing the phase angle, $\theta = \omega t$. Thus, in our
        numerical study, we visualize the attractors in the subspace along with
        their bifurcations by exploring the dynamics in the Poincar\'e
        cross section defined by
\begin{equation}
        \sum = \left\{ (x,dx/dt,y,dy/dt,\theta =\theta _{0}) \in \mathbb{R}^{4}\times S^{1}\right\}\nonumber
\end{equation}
        where $\theta _{0}$ is a constant determining the location of the Poincar\'e
        cross section on which the coordinates $(x,dx/dt,y,dy/dt)\equiv (x_1,v_1,x_2,v_2)$
        of the attractors are expressed. It is worth noticing that  time-reversal symmetry is broken.  Due to the symmetry of the potential $V(x,y)=V(-x,-y)$, the system is invariant under the following transformation 
\begin{eqnarray}
S: (x_1,v_1,x_2,v_2,t)\rightarrow  (-x_1,-v_1,-x_2,-v_2,t+T/2)\nonumber
\end{eqnarray}
where $T=2\pi/\omega$ is the period of the driving force. Therefore in addition to saddle nodes (sn), period doubling (pd) and Hopf (H), one may expect symmetry breaking (sb). This can be evidenced with the eigenvalues spectra obtained from a simple linear stability analysis. Additional features may arise not only because of the driving force but also because the two coupled oscillators are subject to different potentials.

\begin{figure}[h]
\vskip 0.25cm
\includegraphics[width=7cm,height=8cm]{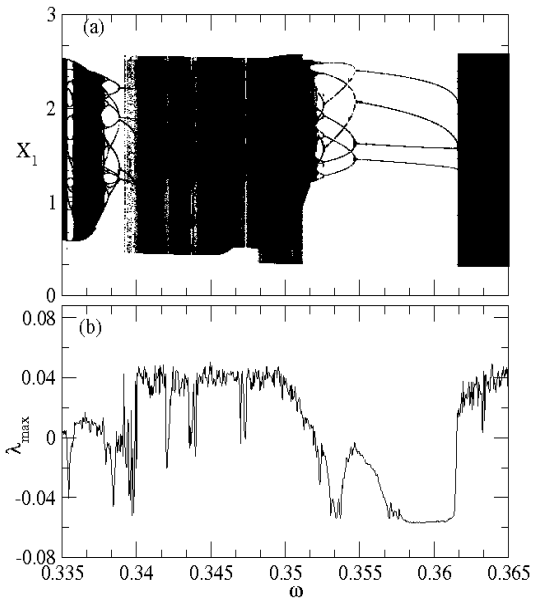}
\caption{\label{figure2}
(a) Bifurcation diagram for $f= 4.0$ and $k=0.1$
showing a sequence of reverse pd route to chaos. Periodic windows sandwitched by chaotic domains are also visible. (b) Lyapunov spectrum corresponding to (a).}
\end{figure}

\section{Bifurcation Structures}
        In the numerical results that follow, we investigate the dependence of the system behaviour at given driving strength $f$ and coupling strength $k$, for varying angular frequency $\omega$. The bifurcation diagrams show the projection of the attractors in the
        Poincar\'e section onto one of the system coordinates versus the control parameter. 
To gain further insight about
        the dynamics of the system under investigation, we compute the Poincar\'e
        sections and Lyapunov spectra $\lambda _{max}$.
 These results are obtained solving eq.(~\ref{equationsmotion}) with the help of the standard fourth-order Runge Kutta algorithm. Starting from the initial condition $(x_1,v_1,x_2,v_2)=(0.0,0.5,0.0,0.5)$, the system is numerically  integrated for 100 periods of the driving force until the transient has died out. To ascertain periodic, quasiperodic and chaotic
        trajectories, the system is further integrated for 180 periods. Next our results are presented in weak, moderate and strong coupling regimes.
\subsection{Weak coupling  ($0<k<1$)}
        At weak coupling strength, we find that the
        bifurcation structures are essentially similar to those commonly found
        in one-dimensional Duffing oscillators. For $k=0.1$, we computed bifurcation diagrams in a large range  of $f$. In Fig.~\ref{figure2}, for instance,
        we set $f= 4.0$ and plot a typical bifurcation diagram in this regime, together with its Lyapunov spectrum $\lambda_{max}$. In the periodic window $0.3375 \le \omega \le 0.34$,
        a sequence of reverse period-doubling (pd) bifurcations yields a period-5 attractor which is later destroyed  at $\omega=0.34$ in a crisis event, say sudden chaos. After the
        large chaotic domain intermingled with windows made up of periodic orbits, another reverse pd cascade
        occurs around $0.352 < \omega < 0.355$. Again, the cascade generates a period-4 attractor which is subsequently destroyed in another crisis event. This bifurcation scenario is in fact the common feature
        of the system as $\omega $ is further increased. However, for $\omega > 0.94$,
        we find periodic orbits dominating the system behaviour. The transitions
        are well characterized by the Lyapunov spectrum shown in Fig.~\ref{figure2}(b).

\begin{figure}[h]
\vskip 0.25cm
\includegraphics[width=7cm,height=5cm]{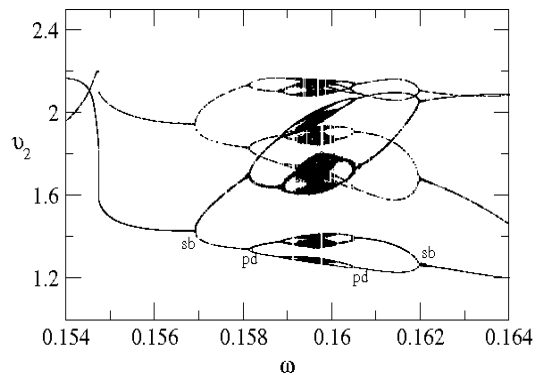}
\caption{\label{figure3}
Bifurcation diagram for $k=0.1$ and $f=4.2$. Bubbles of bifurcation (at $\omega\simeq 0.16$) are sandwiched by period doubling (pd), born from symmetry-breaking (sb).}
 \end{figure}
        As the driving-force amplitude  $f$ increases, different transition
        processes can be captured besides the pd sequence already observed.
        The bifurcation diagram of Fig.~\ref{figure3}, plotted for $f=4.2$,  shows the occurrence of the attractor bubblings (bubbles) sandwiched by symmetry-breaking (sb). The sequence observed, 
consisting of a symmetry breaking, a period-doubling, bubbles, a reversed period doubling and a symmetry breaking bifurcations, can simply be denoted by -sb-pd-(bubbles)-pd-sb-. This means at the first sb point, a symmetrical periodic orbit splits into two coexisting asymmetrical periodic orbits (one being the mirror image of the other); each of the two asymmetrical periodic orbits give rise to another asymmetrical orbit with double period (pd) followed by bubbles. The bubbles  in return generate asymmetrical periodic orbits which join by a pair (reversed pd) to form a symmetrical orbit at the second sb point. 
The sequence -sb-pd-(bubbles)-pd-sb- which is also observed for $f\ge 4.2$, was not reported in previous literature for coupled Duffing oscillators. However, we found a similar sequence in a recent study for coupled ratchets exhibiting synchronized dynamics \cite{vincent2005a}. Subsequently, we will refer to the integer $n$ as the period of an orbit.

\begin{figure}[h]
\vskip 0.25cm
\includegraphics[width=8cm,height=10cm]{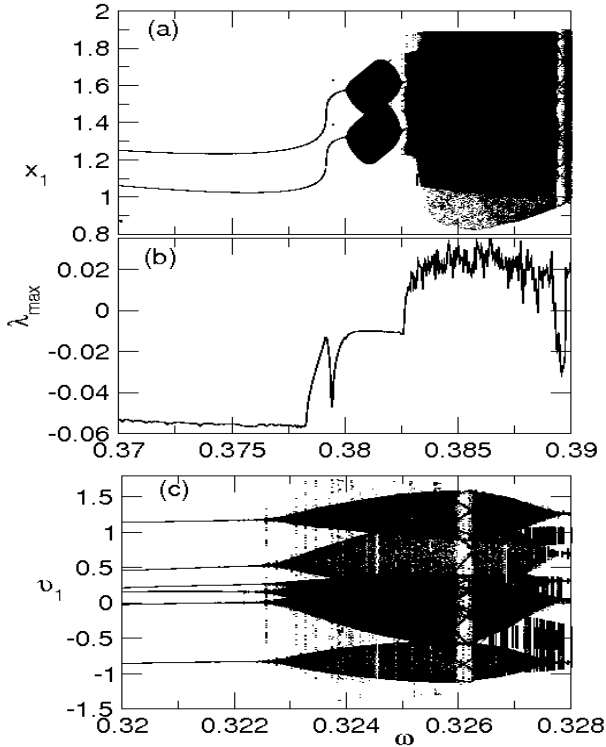}
\caption{\label{figure4} Bifurcation diagram showing varieties of Hopf Bifurcations
for $k=1.0$: (a) $n=2$, $f=1$ and (c) $n=6$, $f=15.7$; (b) Lyapunov spectrum characterizing  (a). Here $n$ represents the period of the periodic orbit.}
\end{figure}

\begin{figure}[h]
\vskip 0.25cm
\includegraphics[width=7cm,height=10cm]{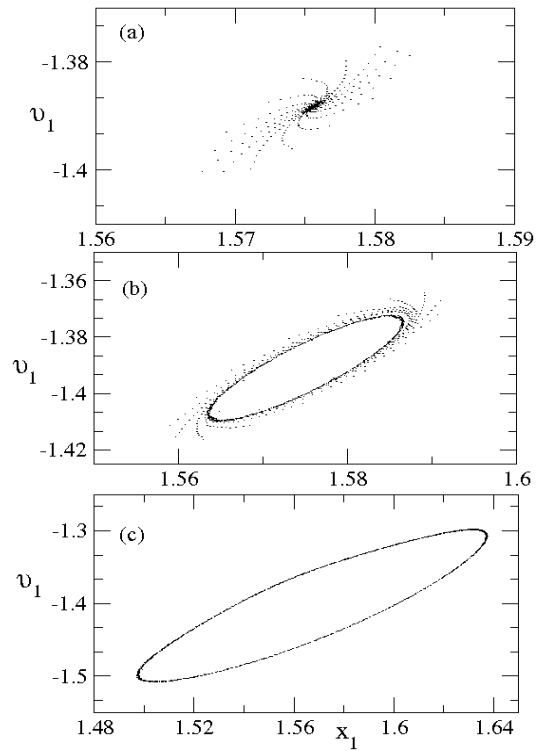}
\caption{\label{figure5}
Poincar\'e sections ($x_1,v_1$) illustrating the supercritical Neimark bifurcation for $k=1$ and $f=1$: (a) $ \omega = 0.38005$, (b) $\omega = 0.38009$ and (c) $\omega = 0.3810$.}
\end{figure}

\begin{figure}[h]
\vskip 0.25cm
\includegraphics[width=7cm,height=8cm]{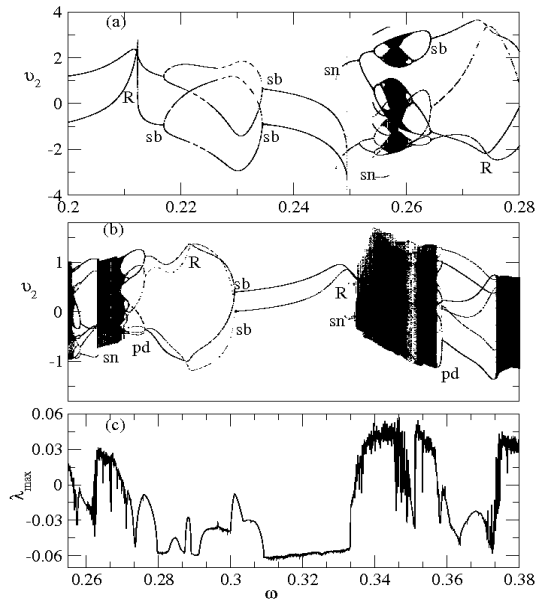}
\caption{\label{figure6}  Bifurcation diagrams for $k=5$, (a) $f=20$ and (b) $f=4$.  Bubbles at around $\omega=0.26$ and several regions of R, pd, sn, sb, and chaos are clearly seen; (c) the Lyapunov spectrum characterizing the bifurcations of (b).}
\end{figure}

\subsection{Moderate coupling ($k=1$)}
        Setting $k=1$, we find (beside pd and sb) an abundance of Hopf bifurcations. In the earlier work~\cite{kozlowski1995,kenfack2003},  no periodic orbits with periods lager than $n=3$ were found to undergo Hopf bifurcations. However in the present model, this number is much  larger and can reach $n=6$. In Fig.~\ref{figure4}, we plot two bifurcation diagrams for (a) $f=1$ and (c) $f=15.7$ in order to illustrate these observations. Hopf bifurcations with orbits of periods (a) $n=2$ and (c) $n=6$ can clearly be seen. Different transitions observed in (a) are well characterized by its Lyapunov spectrum displayed in (b).
By means of Poincar\'e cross-sections, we further zoom structures
in this regime where Hopf bifurcations are predominant. Reported in
Fig.~\ref{figure5} is another bifurcation scenario found in our model, namely the Neimark bifurcation. 
Fig.~\ref{figure5}(a) displays an unstable fixed point spiraling out for $\omega=0.38005$. Further increasing $\omega$, the system moves outward from that unstable fixed point, passing through states like the
one presented in Fig.~\ref{figure5}(b) for $\omega=0.38009$, towards the
attracting invariant closed curve (torus) shown in Fig.~\ref{figure5}(c) for
$\omega=0.381$. This transition is termed secondary Hopf bifuraction or Neimark
Bifurcation~\cite{thompson1986,venkatesan1997}. Moreover, it is a supercritical Neimark bifurcation since the system moves outward from near an unstable fixed point towards the attracting invariant closed curve. In general, the bifurcated solution can be stable, say
supercritical or subcritical otherwise.

\subsection{Strong coupling ($k=5$)}
\begin{figure}[h]
\vskip 0.25cm
\includegraphics[width=7cm,height=9cm]{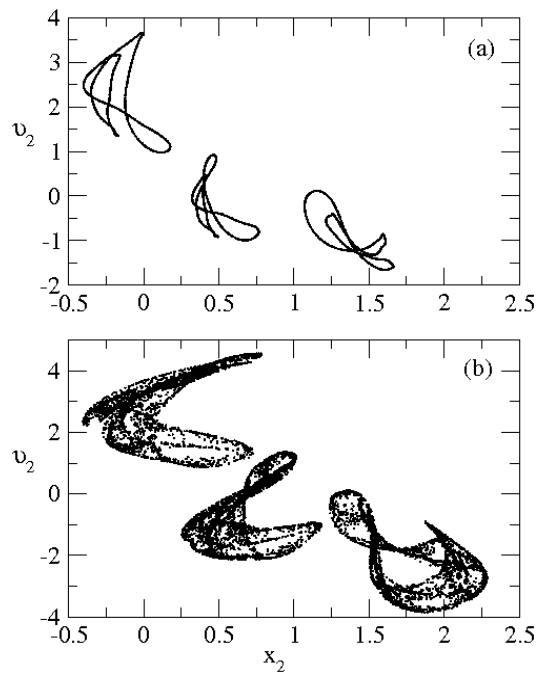}
\caption{\label{figure7}  Poincar\'e sections ($x_2,v_2$) showing the tori-breakdown route to a strange non-chaotic attractor; (a) attractor (folded tori)  $\omega = 1.38$, (b) a strange non-chaotic attractor (distroyed tori) $\omega = 1.412, \lambda _{max} = -0.0174$.}
\end{figure}

Considering $k=5$ as used in ref.~\cite{kozlowski1995,kenfack2003}, and for low $\omega$ values, we
find that periodic orbits dominate for large values of the driving amplitude ($f>20$). However in Fig.~\ref{figure6},
phenomena such as period-doubling (pd), symmetry-breaking
(sb), saddle nodes (sn), resonance (R), bubbles and sudden chaos  can be
found in the frequency range $0.1 \le \omega \le 0.4$, for moderate values of $f\le20$. Remarkably, the Lyapunov exponent characterizing these bifurcations 
exhibits large spikes at bifurcation points. We have observed throughout a variety of Hopf bifurcations
of higher period $n\leq 6$ (not shown), much like in the case $k=1$. Interestingly, another transition found in this model is the passing from low-dimensional to high-dimensional quasiperiodic orbits via tori-breakdown. The sequence of the transformation
from tori (low-dimensional quasiperiodic orbit) to a strange
non-chaotic attractor (high-dimensional quasiperiodic orbit) are shown in
Fig.~\ref{figure7}. In Fig.~\ref{figure7}(a), we plot a poincar\'e section for $\omega = 1.38$,
showing a smooth quasiperiodic attractor visiting three folded tori in the $(x_{2},v_{2})$
sub-space. This folding state of tori is the
onset of tori-breakdown. Upon increasing the frequency to  $\omega = 1.412$, for
instance, the smooth folded tori become completely broken, as can be seen in Fig.~\ref{figure7}(b).
The corresponding Lyapunov exponent, $\lambda _{max} = -0.0174$, shows that the attractor is not chaotic as the {\it eye-test} might tend to show; but rather it is a strange non-chaotic attractor. One would expect chaos from that transition as was reported by Kovlov~\cite{kozlov1999} in a system of coupled Van der Pol oscillators.

\section{Concluding Remarks}
In summary, we have introduced a model consisting of a periodically driven double-well Duffing oscillator linearly coupled to a single-well Duffing oscillator. We have shown that the model admits a very complex dynamical structure that strongly depends on the coupling strength. In addition to the variety of phenomena earlier reported in the coupled Duffing oscillators with identical potentials~\cite{kozlowski1995,kenfack2003}, our model exhibits several other interesting features which are basically attributed to the difference between the two oscillators's potentials. These are essentially attractors bubbling (bubbles), chains of symmetry-breaking bifurcations, Hopf bifurcations born form higher periodic orbits, supercritical Neimark bifurcations and the tori-breakdown route to a strange non-chaotic attractor.  To visualise periodic, quasiperiodic and chaotic attractors of the system, Poincar\'e cross-sections were utilized. Besides, the Lyapunov spectra used to characterise these transitions present spikes at the bifurcation points reflecting a sudden change in the system. In ratchet-like models, such transitions might lead to dramatic changes in transport properties of the system such as current reversals~\cite{mateos2000}. We hope these results would sufficiently complement previous ones and provide a more general overview,  as far as bifurcations structures are concerned, in two coupled driven Duffing oscillators. Finally, it would be interesting to examine broadly a situation wherein the external forcing acts simultaneously on the two oscillators.
\section{acknowledgements}
Authors would like to thank Kamal P. Singh and T. Bartsch for illuminating discussions and for critically reading the manuscript. A. K. acknowledges the Reimar L\"ust grant and the financial support by the Alexander von Humboldt foundation in Bonn.

\end{document}